\documentclass[twocolumn,aps,prl,superscriptaddress]{revtex4}
\usepackage{graphicx}

\begin{document}
\pagestyle{empty} 
\title{Influence of surface roughness on superhydrophobicity}

\author{C. Yang}
\affiliation{IFF, FZ-J\"ulich, 52425 J\"ulich, Germany}
\affiliation{International Center for Theoretical Physics(ICTP),
             I-34014 Trieste, Italy}

\author{U. Tartaglino}
\affiliation{IFF, FZ-J\"ulich, 52425 J\"ulich, Germany}
\affiliation{Democritos National Simulation Center, Via Beirut 2, 34014 Trieste,
 Italy}

\author{B.N.J. Persson}
\affiliation{IFF, FZ-J\"ulich, 52425 J\"ulich, Germany}

\begin{abstract}

Superhydrophobic surfaces, with liquid contact angle $\theta$ greater than
$150^{\circ}$, have important practical applications ranging
from self-cleaning window glasses, paints, and fabrics to low-friction
surfaces. Many biological surfaces, such as the lotus leaf, have 
hierarchically structured surface roughness which is optimized for 
superhydrophobicity through natural selection.
Here we present a molecular dynamics study of liquid droplets in contact with
self-affine fractal surfaces. 
Our results indicate that the contact angle for nanodroplets depends strongly on the 
root-mean-square surface roughness amplitude but is nearly 
independent of the fractal dimension $D_{\rm f}$ of the surface. 

\end{abstract}
\maketitle


\vskip 0.5cm

The fascinating water repellents of many biological surfaces, in particular
plant leaves, have recently attracted great interest for fundamental 
research as well as practical 
applications\cite{Planta,Botany,Dryplant,link,Kaogroup,naturematerial,Mimicking,Lotusleaf}. 
The ability of these surfaces to make water beads off 
completely and thereby wash off contamination very effectively has been
termed the Lotus effect, although it is observed not only on the leaves of
Lotus plant, 
but also on many other plants such as strawberry, raspberry and
so on. 
Water repellents are very important in many industrial and biological
processes, such as prevention of the adhesion of snow, 
rain drops and fog to antennas, 
self-cleaning windows and traffic indicators, low-friction surfaces and
cell mobility \cite{Nakajima, Coulson, Science1}.

Most leaves that exhibit strong hydrophobicity have 
hierarchical surface roughness with micro- and nanostructures made of unwettable
wax crystals.
The roughness enhances the hydrophobic behavior,
so that the water droplets on top tend to become nearly spherical.
As a result the leaves have also a self-cleaning property: the rain drops
roll away removing the contamination particles from the surface,
thanks to the small adhesion energy and the small contact area between
the contaminant and the rough leaf\cite{Planta}. 

The hydrophobicity of solid surfaces is determined by both the chemical 
composition and the geometrical micro- or nanostructure of the 
surface\cite{Rul,Dup,Chen}.
Understanding the wetting of corrugated and porous surfaces
is a problem of long-standing interest in areas ranging from textile science
\cite{Textile} to catalytic reaction engineering\cite{Catalytic}. 
Renewed interest in this problem has been
generated by the discoveries of surfaces with small scale
corrugations that exhibit very large contact angles for water and other 
liquids---in some cases the contact angle is close to $180^\circ$. Such surfaces are
referred to as superhydrophobic\cite{Pearldrops}. 

The contact angle $\theta$ between a flat solid surface and a liquid droplet 
depends on the relation between
the interfacial free energies per unit area: solid/liquid $\gamma_{\rm sl}$,
solid/vapor $\gamma_{\rm sv}$ and liquid/vapor $\gamma_{\rm lv}$.
The Young's equation
$\gamma_{\rm sl}+\gamma_{\rm lv}{\rm cos} \theta =\gamma_{\rm sv},$
results from the minimization of the total free
energy of the system on a flat substrate surface.
Complete wetting corresponds to $\theta=0^{\circ}$, and typically happens on 
solids with high surface energy 
$\gamma_{\rm sv}$.
Liquids on low energy surfaces
tend to form droplets with
high contact angle $\theta$.  

It is well known that the roughness of a 
hydrophobic solid (with $\theta > 90^\circ$ on the flat substrate)
enhances its hydrophobicity. 
If the contact angle of water on such flat solids is of the order
of $100^\circ$ to $120^\circ$, on a rough or microtextured surface it may be
as high as $150^\circ$ to $175^\circ$\cite{Herminghaus,Superstates,Science1}. 
Two distinct models have been proposed to explain this effect. 
The Wenzel model \cite{Wenzel} considers the increase of contact area
due to the surface roughness: this leads to an increase of the
effective free energy of the solid/liquid interface, making the surface more
hydrophobic.
The contact angle $\theta_0$ on the rough surface is obtained from the contact
angle $\theta$ on the microscopically flat surface of the same material
through this equation
\begin{equation}
  \label{Wenzel_equation}
  \cos \theta_0 = r \cos \theta \qquad \mbox{(Wenzel model),}
\end{equation}
where $r=A/A_0$ is the ratio between the real substrate area and the 
nominal (or projected) area $A_0$.

The Cassie model \cite{Cassie} assumes that some air remains trapped between
the drop and the cavities of the rough surface. In this case the
interface free energy $\gamma_{\rm sl}$ must be replaced by a weighted
average of three interface free energies $\gamma_{\rm sl}$, $\gamma_{\rm lv}$
and $\gamma_{\rm sv}$, with the weights depending on the fraction $\phi$
of the area where the contact between the liquid and the solid happens.
The contact angle is given by
\begin{equation}
   \label{Cassie_equation}
  \cos \theta_0 = -1 + \phi (1+\cos \theta)  \qquad \mbox{(Cassie model).}
\end{equation}

Quere states that there exists a critical contact angle $\theta_c$
such that the Cassie state is favored when 
$\theta$ is larger than $\theta_{c}$ \cite{roughideaonwetting}. 
For a micro- or nano structured 
substrate, usually the droplet stays in the Cassie state, but the Cassie state 
can switch (irreversibly) to the Wenzel state when the droplet is pressed against the 
substrate \cite{Carbone}.
The Wenzel droplets are highly pinned, and the transition
from the Cassie to the Wenzel state
results in the loss of the anti-adhesive
properties generally associated with superhydrophobicity.

Many surfaces in nature, e.g., surfaces prepared by fracture (involving crack propagation), 
tend to be nearly self-affine fractal.
Self-affine fractal surfaces have multiscale roughness, sometimes extending from
the lateral size of the surface down to the atomic scale.
A self-affine fractal surface has the property 
that if part of the surface
is magnified, with a magnification which in general is appropriately 
different in the direction perpendicular to the surface as compared to 
the lateral directions, the surface ``looks the same" \cite{P3} 
i.e., the statistical
properties of the surface are invariant under this scale transformation. 

The most important property of a randomly rough surface is the surface 
roughness power spectrum defined as \cite{Nayak, P3, Yang}
\begin{equation}
  \label{powerspectra}
  C(q)=\frac{1}{(2\pi)^2} \int d^2x \ \langle h({\bf x}) h({\bf 0}) \rangle  e^{i {\bf q\cdot x}} 
\end{equation}
Here $h({\bf x})$ is surface height profile and $\langle\cdots\rangle$ 
stands for ensemble average.
We have assumed that the statistical properties of the surface are 
translational invariant and isotropic so that $C(q)$ depends only on the 
magnitude $q=|{\bf q}|$ of the wave-vector $q$.
For a self-affine surface the power spectrum has the 
power-law behavior $C(q) \sim  q^{-2(H+1)}$, where the Hurst exponent 
$H$ is related to the fractal dimension $D_{\rm f}=3-H$. 
Of course, for real surfaces
this relation only holds in some finite wave-vector region
$ q_{0} < q < q_{1}$.
Note that in many cases there is roll-off wave-vector $q_{0}$ below which
$C(q)$ is approximately constant. The mean of the square of the roughness 
profile can be obtained directly from $C(q)$ using
$\sigma^2 = \langle h^2 ({\bf x}) \rangle = \int d^2q \ C(q).$

\begin{figure}
\includegraphics[width=0.46\textwidth]{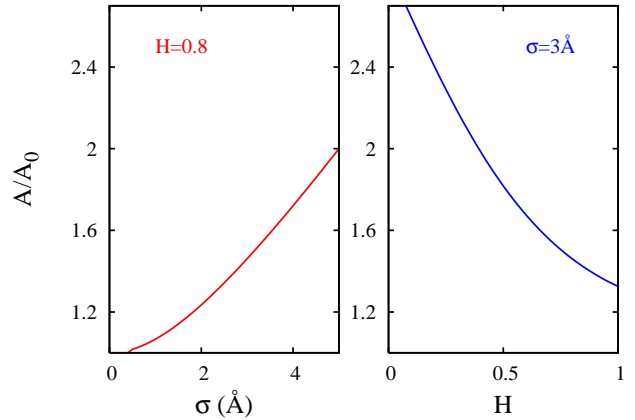} 
\caption{\label{area_ratio}
         The ratio $A/A_0$ between the actual $A$ and the nominal (or projected) $A_0$
         surface area, as a function of the root-mean-square roughness $\sigma$ 
         when Hurst exponent $H=0.8$, and as a 
         function of Hurst exponent $H$ 
         for $\sigma =3$ \AA.}
\end{figure}

For self-affine fractal surfaces $r=A/A_0$ 
is uniquely determined by the
root-mean-square (rms) roughness $\sigma$ and
the fractal dimension $D_{\rm f}$. 
We have\cite{Persson2002}
\begin{equation}
 \label{A/A0}
  A/A_{0}=\int_{0}^{\infty} dx \ (1+x{\xi}^{2})^{1/2} e^{-x}
\end{equation}
where 
${\xi}^2 = \int d^2q \ q^2 C(q)$.
For the surfaces we use in our study, in Fig.~\ref{area_ratio} 
we show the ratio $r=A/A_{0}$ both 
as a function of the root-mean-square roughness $\sigma$,
and as a function of Hurst exponent $H$. 
As expected, $A/A_{0}$ increases with increasing rms-roughness, and decreasing 
Hurst exponent $H$ (or increasing fractal dimension $D_{\rm f}=3-H$). 
Qualitatively, when $D_{\rm f}$ increases at fixed rms-roughness, 
the short-wavelength roughness increases 
while the long-wavelength roughness remains almost unchanged. 

\begin{figure}
\includegraphics[width=0.35\textwidth]{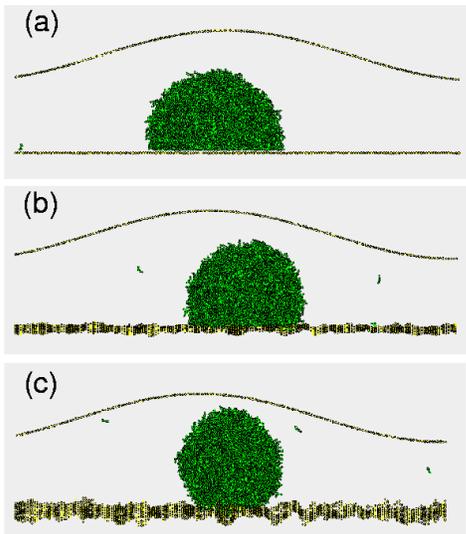} 
\caption{\label{snapshots}
        Snapshots for different root-mean-square roughness. (a) the droplet 
        is in contact with the flat substrate. (b) and (c) are for rough substrates with 
        the root-mean-square amplitude $\sigma =2.3$ \AA\ and
        $\sigma =4.8$ \AA, respectively.}
\end{figure}

We have used Molecular Dynamics calculations to study the
influence of surface roughness on superhydrophobicity.
We have studied hydrocarbon liquid droplets on different self-affine fractal
surfaces. The nano-droplet containes 2364 octane molecules
$\rm{C}_8\rm{H}_{18}$ at $T= 300 \ {\rm K}$, which is between the melting 
and boiling points of octane.
The fractal surfaces were generated as in Ref. \cite{Yang}.
Different fractal surfaces are obtained by changing the root-mean-square
roughness amplitude $\sigma$, and 
the fractal dimension $D_{\rm f}$.
The roll-off wave-vector for the rough surface 
is $q_0=2\pi/L$ with $L=38$ \AA, and the 
magnitude of the short-distance cut-off wave-vector 
$q_1=\pi/a$, where $a= 2.53$ \AA\ is the substrate
lattice constant. 
The (non-contact) cylindrical droplet diameter
is about $104$ \AA, and the size of the 
droplet-substrate contact area varies 
from $\approx 115$ \AA\ (case (a) 
in Fig.~\ref{snapshots}) to
$\approx 60$ \AA\ (case (c)).

\begin{figure}
\includegraphics[width=0.32\textwidth]{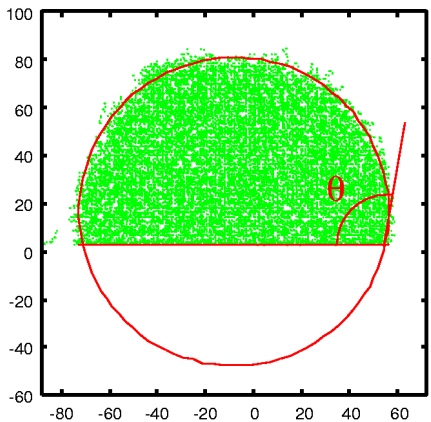} 
\caption{\label{contactangle}
         Determination of the contact angle $\theta$ for the flat
         substrate. Side view.}
\end{figure}

The lubricant molecules are described through the Optimized Potential
for Liquid Simulation (OPLS) \cite{jorgensen1984x1,dysthe2000x1}; this
potential is known to provide density and viscosity of hydrocarbons
close to the experimental one.
We used the Lennard-Jones (L-J) interaction potential between 
droplet atoms and substrate atoms.
The L-J parameters for a hydrophobic surface are chosen such
that the Young contact angle is about $100^{\circ}$ when a droplet sits on 
the flat surface. 
Because of the periodic boundary condition and the size of our system,
the liquid droplet forms a cylinder with the central line 
along the $y$-axis, see Fig.~\ref{snapshots}.
We fit the density profile of the droplet to a cylinder and obtain the contact
angle $\theta=103^{\circ}$ as indicated in Fig.~\ref{contactangle} for
the droplet in contact with the flat substrate.

\begin{figure}
\includegraphics[width=0.4\textwidth]{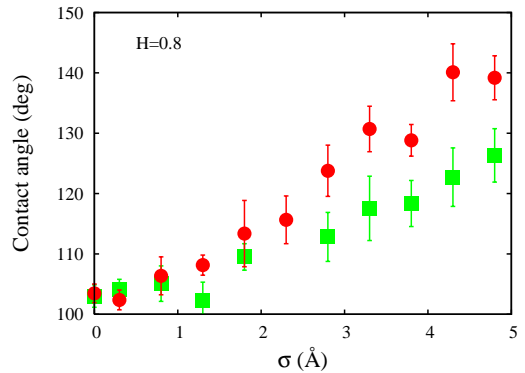} 
\caption{\label{result_rms}
         The contact angle as a function of the root-mean-square
         roughness $\sigma$. The circle points are numerical results from 
         the simulations, while the square points are obtained from the Cassie 
         model (see Eq.~\protect\ref{Cassie_equation}).
         Each data point is an average over several
         snap-shot configurations. 
         The fractal dimension is $D_{\rm f}=2.2$. 
         }
\end{figure}

\begin{figure}
\includegraphics[width=0.4\textwidth]{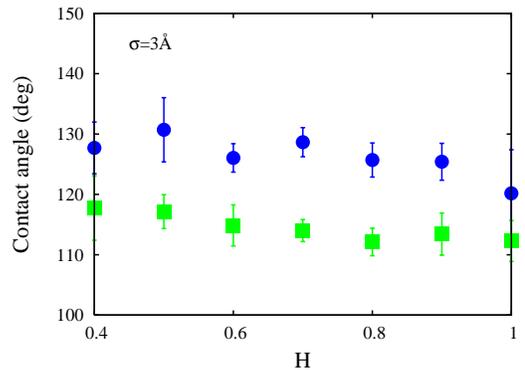} 
\caption{\label{result_H}
         The contact angle $\theta$ as a function of  Hurst exponent $H$ 
         for the rms roughness $\sigma =3$ \AA.
         The circles and squares have the same meaning as that in
         Fig.~\protect\ref{result_rms}
         The fractal dimension is $D_{\rm f}=3-H$.
         }
\end{figure}

The apparent contact angle, $\theta_0$, as a function of the root-mean-square roughness (rms), 
is shown in Fig.~\ref{result_rms} with the fractal dimension $D_{\rm f}=2.2$. 
There is a strong increase in 
$\theta_0$ with
increasing rms-roughness amplitude. 
Fig.~\ref{result_H} shows how $\theta_0$ 
depends on the Hurst exponent $H=3-D_{\rm f}$. 
Note that $\theta_0$ is almost independent of $H$. 

Accordingly to the Wenzel equation, the apparent contact angle $\theta_0$ 
depends only on the surface roughness via 
the ratio $r=A/A_0$. Fig.~\ref{area_ratio} shows that as $H$ decreases from $1$ to
$0.4$ (i.e., $D_{\rm f}$ increases from 2 to 2.6), $A/A_0$ increases by $\sim 50\%$. 
However, the molecular dynamics calculations show that the
apparent contact angle $\theta_0$ is almost
independent of the fractal
dimension, see Fig.~\ref{result_H}. Thus the Wenzel equation cannot be 
used in the present situation. This is consistent with a visual inspection of the liquid-substrate
interface which shows that on the rough substrates,
the droplet is ``riding'' on the asperity top of the substrate, i.e., the
droplet is in the Cassie state. In order to quantitatively verify this,
we have calculated the distances $h(x,y)$ between the bottom
surface of the liquid drop and the rough
substrate surface in the (apparent) contact area. 
From the distribution $P(h)$ of these distances we obtain
the fraction $\psi$ of the (projected) surface area where contact occurs:
$\psi = \int_0^{h_1} dh P(h)$,
where $h_1$ is a cut-off distance to distinguish between contact
and no-contact regions, which has to be comparable to the typical
bond distance (we use $h_1=4$ \AA).
Note that due to the
thermal fluctuations $\psi=\psi_0$ 
for flat surface is less than $1$.
Using the normalized $\phi = \psi/\psi_{0}$, the Cassie model 
predicts the variation of the contact angle
with $\sigma$ and $H$ given in Fig.~\ref{result_rms} and \ref{result_H} (square points).

Fig.~\ref{result_rms} shows that the 
apparent contact angle $\theta_0$ increases strongly with increasing rms-roughness amplitude,
at fixed 
fractal dimension $D_{\rm f}=2.2$,
while it is nearly independent of the fractal dimension $D_{\rm f}$ (see Fig.~\ref{result_H}).
Since increasing the fractal dimension at constant rms roughness amplitude mainly
increases the short-wavelength roughness, we conclude that
the nanoscale wave length
roughness doesn't matter so much in determining the contact angle for 
hydrophobic surfaces, while the long wavelength roughness plays an
important role. We attribute this fact to the strong thermal fluctuations
in the height (or width) $h$ 
of the liquid-solid interface which occur on the nanoscale
even for the flat substrate surface.
Note also that in our model the wall-wall interaction
is long-ranged, decaying effectively as $\sim 1/h^3$, 
so there will be a contribution to the interfacial energy
also for non-contacting surfaces 
which, of course, is not rigorously included in the macroscopic Cassie model.

\begin{figure}
\includegraphics[width=0.4\textwidth]{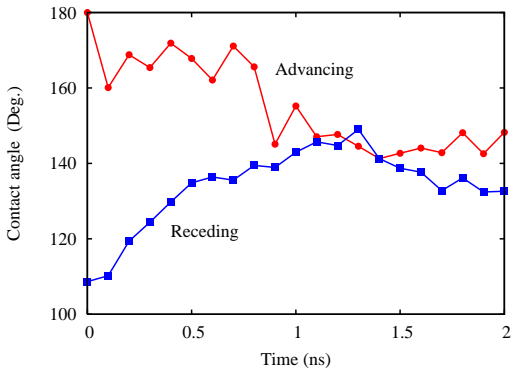} 
\caption{\label{hysteresis}
The Advancing (circles) and receding (squares) contact angle $\theta$ 
as a function of time.
The thermal equilibrium contact angle has been reached
after $\sim 1 \ {\rm ns}$, irrespectively of whether the initial contact angle
is larger or smaller than the equilibrium angle.
         }
\end{figure}

In Fig.~\ref{hysteresis} we study the hysteresis\cite{Chen} in the contact angle $\theta$. 
In one case a spherical droplet was attached to the substrate leading to a decrease
in the contact angle with increasing time (advancing contact angle). 
In the other case the droplet was squeezed into a 
``pancake''-like shape by the upper wall and then released resulting in a contact angle 
which increases with time (receding contact angle). 
In both cases, the thermal equilibrium contact angle has been reached
after $\sim 1 \ {\rm ns}$. Thus, on {\it macroscopic} time scales 
nano-scale roughness 
will not result in any hysteresis in
the contact angle. This is in drastic contrast to simulation studies we
have performed \cite{unpublished} for hydrophilic surfaces, where surface roughness
results in strong pinning of the boundary line; for such surfaces it is therefore 
impossible to study (advancing or receding) droplet contact
angles (as observed on macroscopic time scales) using molecular dynamics. 

Comparing the form of $P(h)$ for the flat
and the most rough surfaces shows that
the system is in the Cassie state, but on the nanoscale the difference between
the Cassie state and the Wenzel state is not so large due to the thermal fluctuations.
This also explain why no hysteresis is observed: The Cassie state is the free energy minimum state
and squeezing the droplet into a pancake shape
does not push the system permanently into the Wenzel state because even if it would go into
this state temporarily, the free energy barrier 
separating the Cassie and Wenzel states is so small that 
thermal fluctuations would almost instantaneously kick it back to the
(free energy minimum) Cassie state.  

In most practical cases it is not possible to modify the surface roughness without simultaneously
affecting the chemical nature of the surface. While this is obvious 
for crystalline materials, where surface
roughening will result in the exposure of new lattice planes with different intrinsic
surface energy, it may also hold for amorphous-like materials, 
where surface roughening may result in a more open
atomic surface structure, with an increased fraction of (weak) unsaturated bonds. 
In our model study a similar effect occurs, and some fraction of the change in the
contact angle with increasing root-mean-square amplitude may be associated with this effect. 
However, the most important result of our study, namely that the 
contact angle is mainly determined by the long-wavelength roughness, should not 
be affected by this fact.
  
To summarize, we have studied the interaction between a liquid
hydrocarbon nano-droplets and
rough surfaces. The macroscopic contact angle $\theta_0$ 
increases with increasing 
root-mean-square roughness amplitude $\sigma$ of the surface, 
but $\theta_0$ is almost unchanged
with increasing fractal dimension $D_{\rm f}$.
There is almost no contact angle hysteresis on the nanoscale. 

This work was partly sponsored by MIUR COFIN No.\ 2003028141-007
MIUR COFIN No.\ 2004028238-002, MIUR FIRB RBAU017S8 R004,
and MIUR FIRB RBAU01LX5H.

\end{document}